\documentclass[%
 reprint,
superscriptaddress,
nofootinbib,
 amsmath,amssymb,
 floatfix,
 aps,
 prl,
]{revtex4-2}
\usepackage{graphicx}% Include figure files

\usepackage{dcolumn}% Align table columns on decimal point
\usepackage{bm}% bold math

\usepackage{color}
\usepackage{braket}
\usepackage[normalem]{ulem}
\usepackage{bbm}

\newcommand{\Tr}{\text{Tr}}

\usepackage[linkcolor={blue},citecolor={blue},colorlinks=true]
{hyperref}
\begin{document}
\title{R\'enyi Mutual Information in Quantum Field Theory}% Force line breaks with \\

\author{Jonah Kudler-Flam}%
 \email{jkudlerflam@ias.edu}
   \affiliation{School of Natural Sciences, Institute for Advanced Study, Princeton, NJ 08540 USA}
  \affiliation{Princeton Center for Theoretical Science, Princeton University, Princeton, NJ 08544, USA}

% \date{\today}

\begin{abstract}
We study a proper definition of R\'enyi mutual information (RMI) in quantum field theory as defined via the Petz R\'enyi relative entropy. Unlike the standard definition, the RMI we compute is a genuine measure of correlations between subsystems, as evidenced by its non-negativity and monotonicity under local operations. Furthermore, the RMI is UV finite and well-defined in the continuum limit. We develop a replica path integral approach for the RMI in quantum field theories and evaluate it explicitly in 1+1D conformal field theory using twist fields. We prove that it bounds connected correlation functions and check our results against exact numerics in the massless free fermion theory.   
\end{abstract}

\maketitle

\textit{Introduction.}---Entanglement is a fundamental feature of quantum systems and has been the subject of extensive research in recent years across a wide range of fields. Studies of the entanglement structure of quantum field theories have proven to be particularly fruitful with landmark achievements including characterizations of critical phases \cite{2003PhRvL..90v7902V}, $c$-theorems \cite{2004PhLB..600..142C}, and spacetime emergence \cite{2006PhRvL..96r1602R} to name a few (see \cite{2019arXiv190708126H, 2022arXiv220113310C} for excellent recent reviews).

In pure quantum states the von Neumann entropy of a subregion $A$, defined as $S_{vN}(A) := -\Tr \rho_A \log \rho_A$, where $\rho_A$ is the reduced density matrix, fully characterizes the entanglement between $A$ and its complement. However, in mixed states, the von Neumann entropy does not provide information regarding correlations between subregions. A useful replacement is the quantum mutual information, which is defined as a linear combination of von Neumann entropies
\begin{align}
    I(A;B) := S_{vN}(A) + S_{vN}(B) - S_{vN}(AB).
    \label{MI_def_eq}
\end{align}
Mutual information characterizes the total (quantum and classical) correlations between subsystems $A$ and $B$ as evidenced by its bound on all connected correlation functions of localized operators \cite{2008PhRvL.100g0502W}
\begin{align}
    I(A;B) \geq \frac{\langle \mathcal{O}_A  \mathcal{O}_B\rangle_c ^2}{2|\mathcal{O}_A|^2_1|\mathcal{O}_B|^2_1}.
    \label{MI_ineq}
\end{align}

A one-parameter family of mutual informations has been studied frequently in the literature by replacing the von Neumann entropies with R\'enyi entropies $S_{\alpha}(A) = \frac{1}{1-\alpha}\log \Tr \rho^{\alpha}$ \cite{2009PhRvL.102q0602F,2009JSMTE..11..001C,2011JSMTE..01..021C,2013JSMTE..02..022H,2010PhRvD..82l6010H,2013JPhA...46B5402C,2016JHEP...02..004H,2015JSMTE..06..021D,2016NatCo...712472D,2013arXiv1303.7221F,2013arXiv1303.6955H,2019JSMTE..09.3107N,2022ScPP...12..117K,2021JHEP...03..146K,2021PhRvR...3c3182K,2020JHEP...01..175K,2018arXiv180909119A,2014PhRvB..90g5132A,2014PhRvB..90d5424S,2011PhRvL.106m5701S,2017PhRvB..96q4201B,2017JPhA...50z4005K,2010arXiv1007.2182M,2013PhRvB..87s5134I,2014PhRvD..89f6015A,2016PhRvL.117v0502L}. While these mutual informations are often easier to compute than \eqref{MI_def_eq}, they are not good measures of correlations because they are not monotonically decreasing under local operations and can even be negative. It is still highly desirable to study generalizations of the mutual information that are well-behaved and provide complementary information to the mutual information. This is in analogy with the fact for pure states that knowing all of the R\'enyi entropies (and hence the full spectrum of $\rho_A$) provides significantly more information about the entanglement structure than simply the von Neumann entropy. In this Letter, we study one such R\'enyi generalization of \eqref{MI_def_eq} that we call the Petz R\'enyi mutual information (PRMI), proving both its utility and computability.

We first note that the mutual information may be written in an equivalent way to \eqref{MI_def_eq} using the relative entropy $D(\rho|| \sigma) := \Tr (\rho \log \rho -\rho\log \sigma)$
\begin{align}
    I(A;B) = D(\rho_{AB}|| \rho_{A}\otimes \rho_B).
\end{align}
The relative entropy is positive semi-definite (for unit trace $\rho$ and $\sigma$) and monotonically decreasing under quantum channels (which include local operations) \cite{lindblad1975completely}. Therefore, the mutual information inherits these qualities. There is a natural one-parameter generalization of the relative entropy which we refer to as the Petz R\'enyi relative entropy (PRRE) defined as
\begin{align}
    D_{\alpha}(\rho||\sigma) := \frac{1}{\alpha-1} \log \left[\Tr\left(\rho^{\alpha} \sigma^{1-\alpha}\right)\right].
\end{align}
The PRRE is positive and limits to the relative entropy when $\alpha \rightarrow 1$. It is furthermore monotonic under quantum channels for $\alpha \in [0,2]$ \cite{lieb1973convex, uhlmann1977relative, petz1986quasi}. We are then motivated to define the PRMI as\footnote{A similar but distinct definition of RMI that replaces $\rho_B$ with a minimization over all states on $B$ has been analyzed in \cite{2013arXiv1310.7028G,2014arXiv1408.3373C,2015JMP....56b2205B,2014arXiv1408.6894H} and given an operational interpretation via quantum hypothesis testing. The PRMI we study provides an upper bound for this RMI. Several interesting results including areas laws for proper R\'enyi generalizations are also given in \cite{2021arXiv210301709S}.}
\begin{align}
    I_{\alpha } (A;B) := D_{\alpha}(\rho_{AB}|| \rho_{A}\otimes \rho_B).
\end{align}
This R\'enyi mutual information is well-behaved and gives new information about the correlation structure between $A$ and $B$. $I_{\alpha}$ is monotonically increasing with $\alpha$, so for all $\alpha \geq 1$, the inequality in \eqref{MI_ineq} continues to hold. A more nontrivial bound can be demonstrated at $\alpha = 1/2$ where the PRMI equals minus the logarithm of Holevo's fidelity \cite{kholevo1972quasiequivalence}. This fidelity satisfies the following bounds
\begin{align}
    {1-e^{-I_{1/2}/2}} \leq T(\rho_{AB}, \rho_A \otimes \rho_B) \leq \sqrt{1-e^{-I_{1/2}}},
\end{align}
where $T(\rho,\sigma) := \frac{1}{2}|\rho-\sigma|_1$ is the trace distance which satisfies
\begin{align}
    \frac{\langle \mathcal{O}_A  \mathcal{O}_B\rangle_c}{2|\mathcal{O}_A|_1|\mathcal{O}_B|_1} \leq T(\rho_{AB}, \rho_A \otimes \rho_B).
\end{align}
Therefore, 
\begin{align}
    I_{1/2}(A;B) \geq \log \left[ \frac{1}{1-\frac{\langle \mathcal{O}_A  \mathcal{O}_B\rangle_c^2}{4|\mathcal{O}_A|^2_1|\mathcal{O}_B|^2_1}}\right].
    \label{Ihalf_bound}
\end{align}
Furthermore, a Pinsker-like inequality holds for the PRREs ($D_{\alpha}\geq 2\min[\alpha,1] T^2$) \cite{2021arXiv210301709S}, such that
\begin{align}
    I_{\alpha} (A;B)\geq \frac{\min[\alpha,1]\langle \mathcal{O}_A  \mathcal{O}_B\rangle_c ^2}{2|\mathcal{O}_A|^2_1|\mathcal{O}_B|^2_1}.
    \label{RMI_ineq}
\end{align}
These inequalities are complementary to the classic result using the usual mutual information \eqref{MI_ineq}. 

We remark that another advantage of relative entropies is that they are well-defined in continuum quantum field theory, unlike the von Neumann entropy which is universally ultraviolet divergent. In particular, the PRREs may be rigorously defined without ever referencing ill-defined density matrices by using Tomita-Takesaki theory \cite{petz1985quasi}. Nevertheless, we will use density matrices due to their convenience in illustration of explicit computations.

\textit{Replica Approach.}---Replica tricks have been leveraged considerably to compute different correlations measures (see e.g.~\cite{1994NuPhB.424..443H, 2004JSMTE..06..002C,2012PhRvL.109m0502C,2014PhRvL.113e1602L,2019PhRvL.122n1602Z}). The main technical contribution of this work is to develop a replica path integral approach for the computation of the PRMI which involves a new ingredient to account for the difference between the two states $\rho_{AB}$ and $\rho_A \otimes \rho_B$. It is generally easier to evaluate traces of positive integers of density matrices. In order to account for the exponent of $1-\alpha$ in the PRRE, we introduce the following ``double'' replica trick that has been used previously in random matrix theory calculations \cite{2021PRXQ....2d0340K}
\begin{align}
     D_{\alpha}(\rho||\sigma) = \lim_{m\rightarrow 1-\alpha}  \frac{1}{\alpha-1}\log \left[\Tr\left(\rho^{\alpha} \sigma^{m}\right)\right].
     \label{PREE_replica}
\end{align}
Here, $\alpha$ and $m$ are first taken to be positive integers. Then the function is analytically continued by taking $m\rightarrow 1-\alpha$ and arbitrary $\alpha$. While the analytic continuation of functions defined on the positive integers is a priori ambiguous, this invaluable approach has become standard and we will see that the natural analytic continuation passes several consistency checks.

\begin{figure}
    \centering
    \includegraphics[width = .48\textwidth]{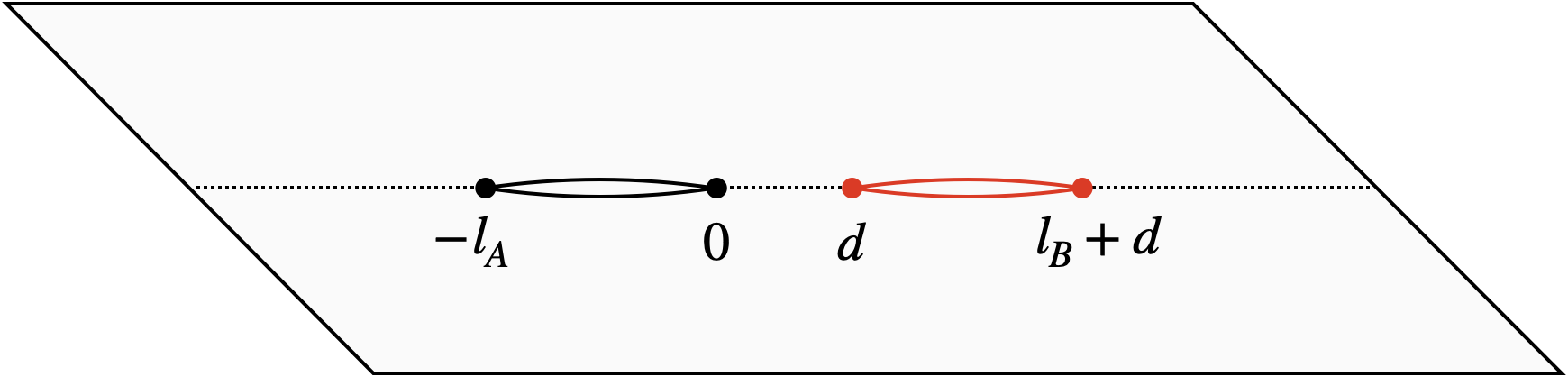}
    \caption{The path integral computing the matrix elements of $\rho_{AB}$ in the vacuum state.}
    \label{path_int_fig}
\end{figure}

\begin{figure}
    \centering
    \includegraphics[width = .48\textwidth]{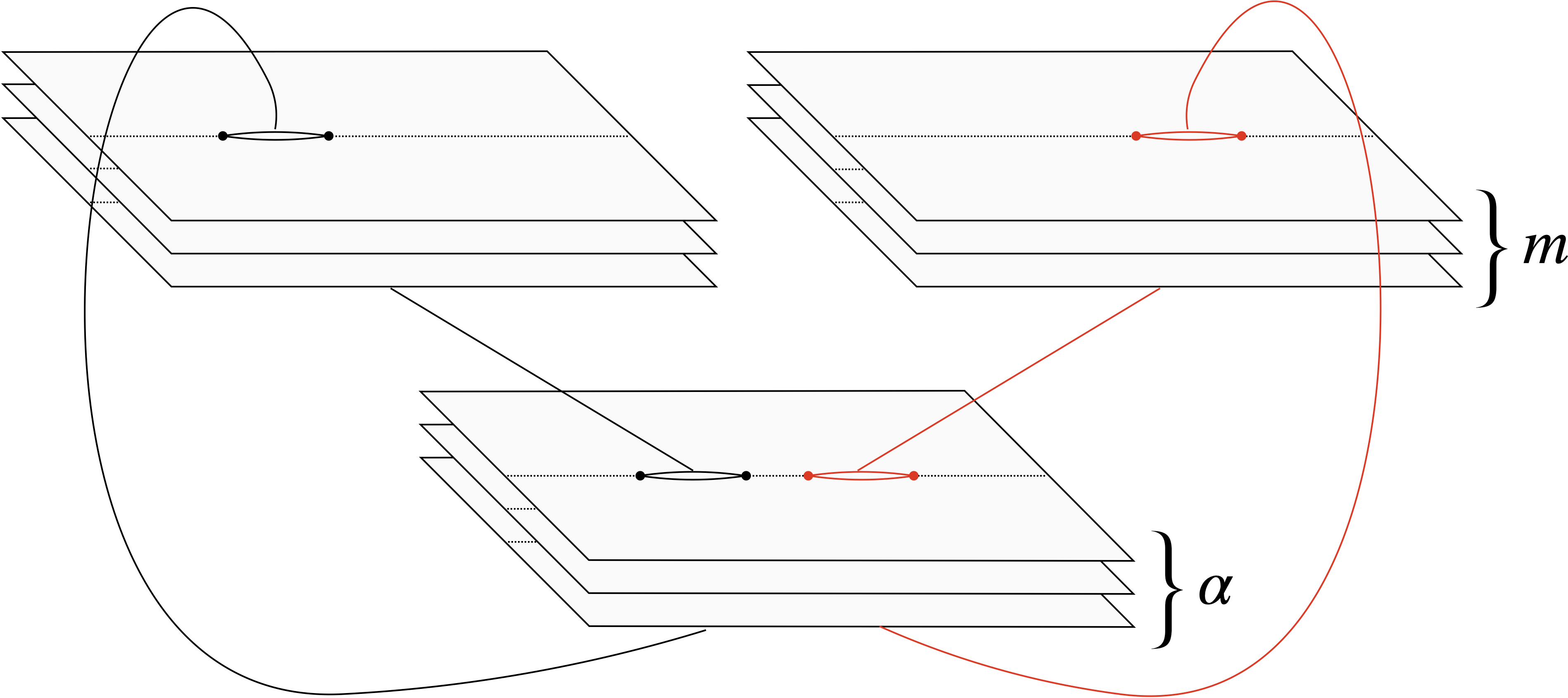}
    \caption{The replica manifold that computes $\Tr (\rho_{AB}^{\alpha}(\rho_A \otimes \rho_B)^m)$. The lines correspond to the gluing structure between the replicas.}
    \label{replica_fig}
\end{figure}

Quantum states and similarly reduced density matrices may be conveniently prepared using path integrals with the matrix elements corresponding to boundary conditions (see e.g.~\cite{2009JPhA...42X4005C}). For concreteness, we now specify to disjoint intervals $A = (-l_A, 0 )$ and $B = (d, l_B+d)$ in 1+1D quantum field theories where pictures are simplified, though we stress that the replica approach is much more general. In the vacuum state, the (unnormalized) density matrix $\rho_{AB}$ corresponds to the Euclidean path integral on the plane with two slits at the locations of the intervals (see Figure \ref{path_int_fig}). Similarly, the density matrices $\rho_A$ and $\rho_B$ are path integrals with a single slit. The tensor product state is simply the product of these two path integrals. The trace structure in \eqref{PREE_replica} enforces a gluing between the $\alpha + 2 m$ replica sheets as shown in Figure \ref{replica_fig}. Note that, unlike the von Neumann entropy replica trick, the gluing between the sheets is different for regions $A$ and $B$. Ordering the sheets such that the $\alpha$ double-slitted sheets come first, the $m$ $A$-slitted sheets next, and the $m$ $B$-slitted sheets last, the permutation group elements on the two regions in cycle notation are
\begin{align}
    g_A &= (1,\dots, \alpha, \alpha + 1,\dots \alpha + m), 
    \\
    g_B &= (1,\dots, \alpha, \alpha + m+1,\dots \alpha + 2m).
\end{align}
Crucially, the composition of $g_A^{-1}$ and $g_B$ does not give the identity element
\begin{align}
    g_{A}^{-1} g_B = (1,\alpha + m, \dots , \alpha+ 1,\alpha+m+1,\dots ,\alpha +2m),
\end{align}
where the first ``$\dots$'' are decreasing and the second are increasing. The partition functions on these replica manifolds can be conveniently evaluated by computing correlation functions of twist operators generating the permutations at the boundaries of $A$ and $B$ in the product theory of $\alpha + 2m$ copies. We now demonstrate this explicitly in 1+1D conformal field theory.

\textit{Conformal Field Theory.}---In 1+1D, the boundaries of $A$ and $B$ are points, so the twist operators are local fields. The conformal dimensions of the twist fields may be determined directly from their cycle structure
\begin{align}
    \Delta := \Delta_{g_A} = \Delta_{g_B} &= \frac{c}{12}\left(\alpha + m -\frac{1}{\alpha + m} \right),
    \\
    \Delta_{g_A^{-1}g_B} &= \frac{c}{12}\left(2m+1 -\frac{1}{2m+1} \right),
\end{align}
where $c$ is the central charge.

The twist fields are scalar conformal primaries, so we are now in position to perform explicit calculations. We begin with a consistency check by considering the pure state limit where $B$ is the complement of $A$. The PRMI of pure states must equal $2 S_{3-2\alpha}(\rho_A)$ as can be easily deduced from the Schmidt decomposition of pure states. In this case
\begin{align}
    \Tr\left(\rho^{\alpha}\sigma^m\right) = \langle \sigma_{g_A^{-1}g_B}(0) \sigma_{g_A^{-1}g_B}^{-1}(l_A)\rangle 
= l_A^{-2\Delta_{g_A^{-1}g_B}},
\end{align}
where we have used in the second equality that the two point function is fixed by conformal invariance. We then find
\begin{align}
    I_{\alpha} = c\frac{2(2-\alpha)}{3(3-2\alpha)}\log \frac{l_A}{\epsilon},
\end{align}
where $\epsilon$ is an ultraviolet cutoff that was implicit in the definition of the twist fields and may be thought of as the lattice spacing.
This precisely matches $2 S_{3-2\alpha}(\rho_A)$ which was computed in \cite{1994NuPhB.424..443H, 2004JSMTE..06..002C}. We observe that the PRMI diverges as $\alpha \rightarrow 3/2$ which is still well within the regime where PRRE is monotonic under local operations. This is because the density matrix has an infinite number of non-zero eigenvalues. When $\alpha \rightarrow 3/2$, the PRMI becomes twice the max entropy (or Hartley entropy) which equals the logarithm of the rank of $\rho_{AB}$. See \cite{2008PhRvA..78c2329C} for an explicit calculation of this continuous eigenvalue spectrum of $\rho_{AB}$ in 1+1D conformal field theories.

After having passed this consistency check, we may progress to new results. For adjacent intervals, we must evaluate the following three-point function, which is also fixed by conformal symmetry
\begin{align}
    \Tr\left(\rho^{\alpha}\sigma^m\right) = \langle \sigma_{g_A}(-l_A) \sigma_{g_A^{-1}g_B}(0)\sigma_{g_B}^{-1}(l_B)\rangle 
    \nonumber
\\
= \frac{C_{\alpha,m}}{(l_Al_B)^{\Delta_{g_A^{-1}g_B}}(l_A+l_B)^{2\Delta-\Delta_{g_A^{-1}g_B}}},
\end{align}
where $C_{\alpha,m}$ is the operator product expansion (OPE) coefficient that takes a universal (only dependent on $c$, $\alpha$, and $m$) form due to the replica manifold being genus-$0$, as can be readily checked using the Riemann-Hurwitz formula. The replica limit leads to
\begin{align}
\label{adj_vac_eq}
    I_{\alpha}(A;B) = c\frac{(2-\alpha)}{3(3-2\alpha)}\log \frac{l_A l_B}{\epsilon(l_A + l_B)} +O(1).
\end{align}
We can also back away from the adjacent intervals limit by considering the four-point function
\begin{align}
        \hspace{-.1cm}\Tr\left(\rho^{\alpha}\sigma^m\right) = \langle \sigma_{g_A}(-l_A) \sigma_{g_A}^{-1}(0)\sigma_{g_B}(d)\sigma_{g_B}^{-1}(l_B+d)\rangle .
\end{align}
This is no longer fixed by conformal symmetry and instead depends on the full operator content of the theory. However, conformal symmetry still buys us the prediction that the PRMI only depends on the conformally invariant cross-ratio
\begin{align}
    x = \frac{l_A l_B}{(l_A+d)(l_B +d)}.
\end{align}
For sufficiently small (but finite) $d$, \eqref{adj_vac_eq} is valid for disjoint intervals by replacing $\epsilon$ with $d$. For sufficiently large $d$, one may take the OPEs of the $A$ operators and $B$ operators separately. The leading term is the identity operator which leads to trivial PRMI as expected for a correlation measure of distant intervals. There will be subleading corrections controlled by the lightest operators in the theory, in analogy with \cite{2011JSMTE..01..021C,2013JPhA...46B5402C}.

We may also study the correlations in the thermal Gibbs state at inverse temperature $\beta$ by considering the correlation function on the cylinder with circumference $\beta$. For adjacent intervals,
\begin{align}
    \Tr\left(\rho^{\alpha}\sigma^m\right) = \langle \sigma_{g_A}(-l_A) \sigma_{g_A^{-1}g_B}(0)\sigma_{g_B}^{-1}(l_B)\rangle_{\beta} .
\end{align}
We can map this correlation function to the complex plane using the conformal transformation $e^{2\pi z/\beta}$. The twist fields transform as Virasoro primaries, such that
\begin{align}
    \Tr\left(\rho^{\alpha}\sigma^m\right) =\left(\frac{2\pi}{\beta} \right)^{\Delta_{g_A^{-1}g_B}} \left(\frac{4\pi^2}{\beta^2} e^{\frac{2\pi}{\beta}(l_B-l_A)}\right)^{\Delta }
    \nonumber
    \\
    \langle \sigma_{g_A}(e^{-2\pi l_A/\beta}) \sigma_{g_A^{-1}g_B}(1)\sigma_{g_B}^{-1}(e^{2\pi l_B/\beta})\rangle_{\mathbb{C}} .
\end{align}
Taking $l_A = l_B = l$, we have
\begin{align}
\label{adj_finiteT_eq}
    I_{\alpha}(A;B) = c\frac{(2-\alpha)}{3(3-2\alpha)} \log \left(\frac{\beta}{2\pi \epsilon} \tanh \frac{\pi l}{\beta}\right)+O(1)
\end{align}
which is monotonically decreasing with temperature, reflecting the fact that finite temperature effects destroy entanglement. This is consistent with the thermal area laws derived in \cite{2021arXiv210301709S}.

\begin{figure}
    \centering
    \includegraphics[width = .48\textwidth]{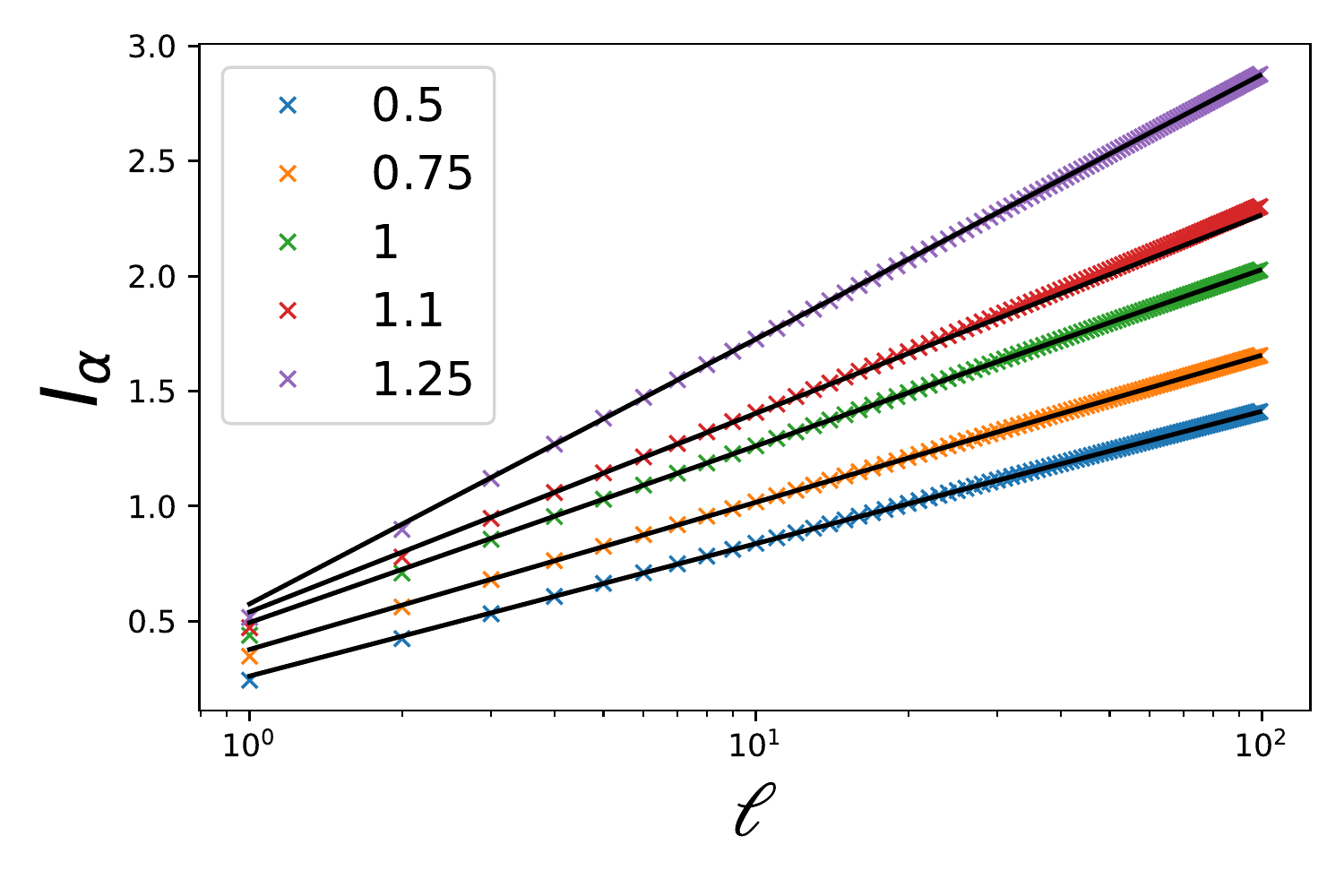}
    \caption{The PRMI is plotted for various values of $\alpha$ which are labeled in the legend. Here, $l = l_A = l_B$. The $\times$'s are numerical data while the solid black curves are \eqref{adj_vac_eq} with the regularization dependent $O(1)$ shift fitted.}
    \label{fig:adjacent}
\end{figure}

\begin{figure}
    \centering
    \includegraphics[width = .48\textwidth]{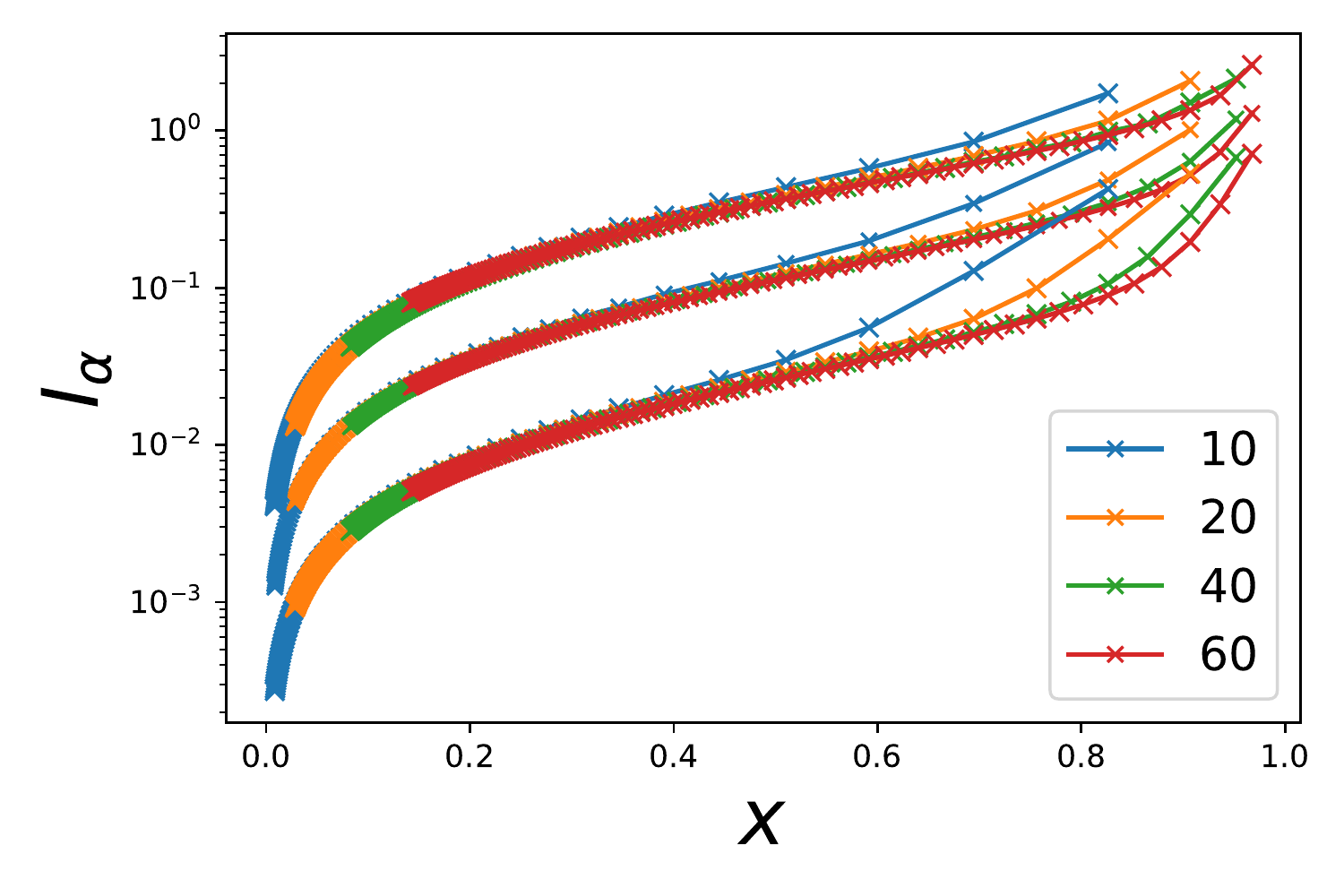}
    \caption{The PRMI is plotted for different values of $l=l_A = l_B$ as labeled in the legend. The three bands are $\alpha = \{.1,.5,1.25 \}$ from bottom to top. The collapsing of the curves show that the PRMI only depends on the conformal cross-ratio, though we see minor deviations when $l$ is not large enough and hence experiences finite-size effects.}
    \label{fig:disjoint}
\end{figure}

\textit{Free Fermions.}---We would like to support our universal conformal field theory results with an independent check. For this purpose we study a single massless chiral fermion with central charge $1/2$ and Hamiltonian $H = -\frac{i}{2} \int dx \left( \psi^{\dagger} \partial \psi - \partial \psi^{\dagger} \psi\right)$. This Hamiltonian may be discretized on the lattice as
\begin{align}
    H = -\frac{i}{2}\sum_{j}\left(\psi^{\dagger}_j\psi_{j+1}-\psi^{\dagger}_{j+1}\psi_{j} \right),
\end{align}
where $\{\psi_i ,\psi_j \} = \delta_{ij}$. The ground state is a so-called Gaussian state, which is fully characterized by the two-point functions $[C]_{jl} = \langle \psi_j \psi_l^{\dagger}\rangle $ \cite{2003JPhA...36L.205P}. Following the techniques of \cite{2014PhRvE..89b2102B, 2018JHEP...09..166C}, it is straightforward to derive an expression for the PRMI in terms of the correlation matrix restricted to regions $A$ and $B$
\begin{align}
    I_{\alpha}(A;B) = -\frac{\alpha\Tr \log \left(1-C \right)}{1-\alpha}-\Tr \log \left(1-C' \right)
    \nonumber
    \\
    -\frac{\Tr \log \left( 1+ \left(\frac{C}{1-C}\right)^{\frac{\alpha}{2}}\left(\frac{C'}{1-C'}\right)^{1-{\alpha}{}}\left(\frac{C}{1-C}\right)^{\frac{\alpha}{2}} \right)}{1-\alpha}.
\end{align}
Here
\begin{align}
    C' = \begin{pmatrix}
    C_{AA} & 0
    \\
    0 & C_{BB}
    \end{pmatrix},
\end{align}
where $C_{AA}$ and $C_{BB}$ are submatrices of $C$. The setting of the off-diagonal terms to zero reflects that there are no correlations between $A$ and $B$ in $C'$. For the ground state, the correlation function is given by \cite{2020JHEP...05..103B}
\begin{align}
    [C]_{jl} = \begin{cases}
    \frac{(-1)^{j-l}-1}{2\pi i (j-l)} ,  & j \neq l
    \\
    \frac{1}{2}, & j = l
    \end{cases}.
\end{align}
We may now efficiently simulate the PRMI in the ground state. In Figure \ref{fig:adjacent}, the PRMI is shown for various values of $\alpha$ with excellent agreement with the theoretical prediction \eqref{adj_vac_eq}. An additional advantage of numerics is that we may reliably move beyond the universal adjacent intervals limit. We demonstrate in Figure \ref{fig:disjoint} that the PRMI for disjoint intervals only depends on the conformally invariant cross-ratio, as expected. 

\textit{Discussion.}---In this Letter, we have developed a replica path integral method for computing a R\'enyi generalization of the quantum mutual information. This RMI was shown to obey the important properties of monotonicity under local operations and positivity. Furthermore, we demonstrated that it places strong bounds on connected correlation functions. In 1+1D conformal field theory, we found that the PRMI behaves universally for a variety of set-ups, confirming this in the chiral fermion conformal field theory.

There are many (known and unknown) applications of the techniques we developed. For instance, it would be fascinating to study the dynamics of the PRMI, its behavior in typical (Page) states and tensor networks, and develop a holographic formula. It may also be generalized to other families of RMIs using different divergences such as the sandwiched R\'enyi relative entropy \cite{2014CMaPh.331..593W,2013JMP....54l2203M}. Each RMI has their own operational interpretation and will sharpen our understanding of correlations in many-body quantum states. One may furthermore hope the RMIs to impose nontrivial constraints on correlation functions beyond \eqref{Ihalf_bound} in the spirit of \cite{2019JHEP...01..059L}. We hope to present progress in some of these directions in upcoming work \cite{WIP}.

\begin{acknowledgments}
\textit{Acknowledgments.}---I thank Alvaro Alhambra, Pasquale Calabrese, Nima Lashkari, and Mark Wilde for discussions and comments and Laimei Nie, and Akash Vijay for related collaboration. I thank Viktor Eisler for pointing out typos in \eqref{adj_vac_eq} and \eqref{adj_finiteT_eq} in an earlier version. I am supported by the Institute for Advanced Study and the National Science Foundation under Grant No.~PHY-2207584.
\end{acknowledgments}

\bibliography{main}

\end{document}